# Mental Wormholes and the Inner Workings of Creative Thought

Liane Gabora
University of British Columbia

Please address correspondence regarding this article to:
Liane Gabora
Department of Psychology, University of British Columbia
Okanagan Campus, Kelowna BC, Canada, V1V 1V7
Email: liane.gabora@ubc.ca
Tel: 604-358-7821 / 250-807-9849

## Introduction

Most creative outputs are readily classified as belonging to a particular domain such as art, music, or technology. But does that mean the creative thinking that goes into creative projects respects these conventional domain boundaries? This question provides the departure point for this chapter. It is an important question because the answer has implications for how the creative process works, as well as for computational models of creativity, and thus it provides the departure point for this chapter. Next we look at how the creator sifts out from its vast contents those items to incorporate into a creative project, and converts these raw materials into a form that "gels" with the project. We then explore the hypothesis that the creative mind forges "mental wormholes" that connect concepts and percepts, often from seemingly disparate domains. Finally, we take a brief look at the strangely non-compositional interactions that take place in these mental wormholes, and techniques being advanced to model and understand these interactions.

## Does Creative Thought Respect Domain Boundaries?

I remember as a child placing everything in the world into categories such as "things that don't change" (e.g., a rock), "things that go back and forth between two possibilities" (e.g., a metronome), "things that circle around to their starting point" (e.g., seasons, or music that always goes back to the chorus), things that spiral around to a slightly different starting point" (e.g., music that is transposed to a different key before returning to the chorus, or a story in which the character changes before returning home), and so forth. Whether a pattern was implemented as musical notes versus words on a page seemed superficial compared to the fact that they were both instances of, say, "spiral". Amazingly enough, the adults in my life didn't see this as a profound and useful way of carving up reality.

Of course, reality doesn't *come* carved up into domains. However, since we all come into this world equipped with sensory apparatuses that parse information in specific ways, and we are educated in ways that reinforce the categorization of knowledge into specific domains (mathematics, music, art, and so forth), it is natural that we come to accept these domains as real, or at least as pragmatically useful social constructs. But that doesn't mean that they necessarily reflect how we think. There is evidence that people do not think in ways that respect conventional domain boundaries (e.g., musical thought is not restricted to concepts such as rhythm and timbre) . Ingredients for creative thinking such as remote associates (e.g., "cottage" and "cheese") (Mednick, 1962), analogies (Gentner, 1983), 'seed incidents' (e.g., falling in love inspires a song) (Doyle, 1998), and inspirational sources often come from *beyond* the problem domain. For example, in one study, 66 creative individuals in a variety of disciplines were asked to list as many influences on their creative work as they could (Gabora & Carbert, 2015). Rather astonishingly, not only were cross-domain influences found to exist, they more widespread than within-domain influences (Table 1). Moreover, this was the case even when not just narrow within-domain influences (e.g., painter influenced by paintings) but also broad within-domain influences (e.g., painter influenced by Spirograph toy) are classified as "within-domain influences".

Table 1: Number of participants in each creative domain (N), and the raw number (r) and percentage (%) of influences that were cross-domain (CD), within-domain: narrow (WD-n), within-domain: broad (WD-b, and uncertain (U). Percentages are in brackets. A dash indicates that no examples were present in the data. (From Gabora & Carbert, 2015).

| Creative | N | CD | | WD-n | | WD-b | | U | |
|---|---|---|---|---|---|---|---|---|---|
| Domain | | r | (%) | r | (%) | r | (%) | r | (%) |
| Painting | 44 | 21 | (48) | 12 | (27) | 4 | (9) | 6 | (14) |
| Drawing | 8 | 2 | (25) | 2 | (25) | – | – | 3 | (38) |
| Photography | 4 | 2 | (50) | – | – | 1 | (25) | – | – |
| Sculpture | 5 | 3 | (60) | – | – | – | – | 1 | (20) |
| Music | 3 | 1 | (33) | 2 | (68) | – | – | 1 | (33) |
| Writing | 2 | 2 | (100) | 1 | (50) | – | – | 1 | (50) |
| TOTAL | 66 | 31 | **(47)** | 17 | **(27)** | 5 | **(8)** | 12 | **(18)** |

These results suggest that even when a particular creative output is categorized as belonging to a single domain (e.g., a painting belongs to the domain of art), the thinking that went into it may well have spanned many domains. In fact, analysis of the data suggested that just about anything encoded in memory can potentially be recruited into a creative thought process.

So a first point to be made about the inner workings of creative thought is that it can incorporate ingredients from beyond the domain in which it will eventually be expressed. This leads to two questions. First, how does the creator sift out from its vast contents those items to incorporate into a given creative thought process? Second, how does it convert this raw material into a form that is amenable to re-expression in a different domain?

## Aware-Restructure-Express (ARE)

Possible answers to these questions are suggested by a number of theories of creativity, including the *honing theory of creativity* (Carbert, Gabora, Schwartz, & Ranjan, 2014; Gabora, 2005, 2010a, in prep.). The basic idea is as follows. The creative mind is perpetually on the lookout for arenas of creative potential or incomplete understanding that Torrance (1962) refers to as the *gap*. The gap is arousal-inducing (where arousal may be either positively or negatively valenced). The creative process involves sequentially considering this gap from different perspectives, at different levels of detail, such that arousal dissipates. The new perspectives (1) restructure information (radically or imperceptibly) in the region of the gap, and (2) suggests a subsequent perspective to look at it from. The finding of new perspectives may be facilitated by shifting to an associative mode of thought, such that cell assemblies that respond to not just central but also peripheral features of the task are recruited. Since these cell assemblies may previously have encoded memories related to the task but in an atypical way, they can be the gateway to creative solutions (Gabora, 2002, 2010a). Eventually the new perspectives have restructured task-related ideas to the extent that they exhibit an acceptable level of consistency (Gabora, 1999), arousal decreases, and the creative outcome is expressed.

We can refer to this process of becoming aware of an arousal-inducing gap, sequentially restructuring it, and expressing it in restructured form, as the Aware-Restructure-Express process, or ARE. Note that domain-specific operations are *in service of* ARE; for example, one goes through the motions of committing paint to a canvas or words to page due to the arousing effect of the sense of incomplete understanding or creative potential. Findings that creativity can be intrinsically rewarding (Gruber, 1995; Kounios & Beeman, 2014; Martindale, 1984), and therapeutic (Barron, 1963; Forgeard, 2013), and that high levels of creativity are correlated with positive affect (Hennessey & Amabile, 2010) suggest that one may be further motivated by anticipation of the pleasant and perhaps even therapeutic impact of the creative process.

## Implications for Whether Creativity is Domain-Specific or Domain-General

The apparent domain-specificity of creativity—the fact that those who are eminently creative in one domain are unlikely to be eminently creative in another—may be an artifact of our focus on the external outcome of creative thinking, i.e., the product. When we focus instead on the internally transformative effects of creativity it appears to be much more domain-general; many if not most individuals have multiple avenues for transformative self-expression. This is particularly the case when we consider the creative aspects of activities that are not prototypically creative, such as organizing parties, fixing household appliances, or even, deceiving others.

There is evidence that when people express themselves creatively in different domains their creative outputs bear the same recognizable distinctive style or 'voice' (Gabora, 2010; Gabora, O'Connor, & Ranjan, 2012; Ranjan, 2014). In one study, art students were able to identify at significantly above-chance levels which famous artists created pieces of art they had not seen before, as well as which of their classmates created pieces of art they had not seen before. More surprisingly, art students also identified the creators of non-painting artworks that they had not seen before. Similarly, creative writing students were able to identify at significantly above-chance levels passages of text written by famous writers that they had not encountered before, and passages of text written by their classmates that they had not encountered before. Perhaps most surprising, creative writing students also identified at significantly above-chance levels which of their classmates created a work of art, i.e., a creative work in a domain other than writing. The finding that creative style is recognizable across domains is incompatible with the view that creativity is domain-specific.

Not only does a creator's personal style come through in different domains, but an inspirational source leaves a recognizable trace on a creative work in a different domain. In a study in which pieces of music were re-interpreted as paintings, naïve participants were able to correctly identify at significantly above chance which piece of music inspired which painting (Ranjan, Gabora, & O'Connor, 2014; Ranjan, 2014). Although the medium of expression is different, something of its essence remains sufficiently intact for an observer to detect a resemblance between the new work and the source that inspired it. These results lend empirical support to the largely anecdotal body of evidence that cross-domain influence is a genuine phenomenon, and suggested that, at their core, creative ideas are less domain-dependent than they are generally assumed to be.

## Mental Wormholes

Creativity is often said to involve forging new associations amongst concepts and percepts (Ward, Smith, & Vaid, 1997). Thus percepts and concepts are the building blocks from which creative ideas are generated. A *percept* is a mental impression of something perceived by the senses, such as the sight of a particular kitchen. We often interpret percepts in terms of related percepts we have encountered in the past, i.e., we spontaneously categorize a particular kitchen as an instance of the concept KITCHEN. A *concept* is a mental construct that enables us to interpret the present situation in terms of similar previous situations. Concepts can be concrete, like KITCHEN, or abstract, like VIRTUE. Although concepts have traditionally been viewed as internal structures that *represent* a class of entities in the world, increasingly they are thought to have no fixed representational structure, their structure being dynamically influenced by the contexts in which they arise (Hampton, 1997). For example, you might consider the concept KITCHEN in the context of a doll house, in which case it is tiny.

We can use the tem *mental wormholes* to refer to interactions between percepts and concepts from different domains, as schematically illustrated in Figure One. Since these interactions lie at the heart of the creative act, let us examine them more closely.

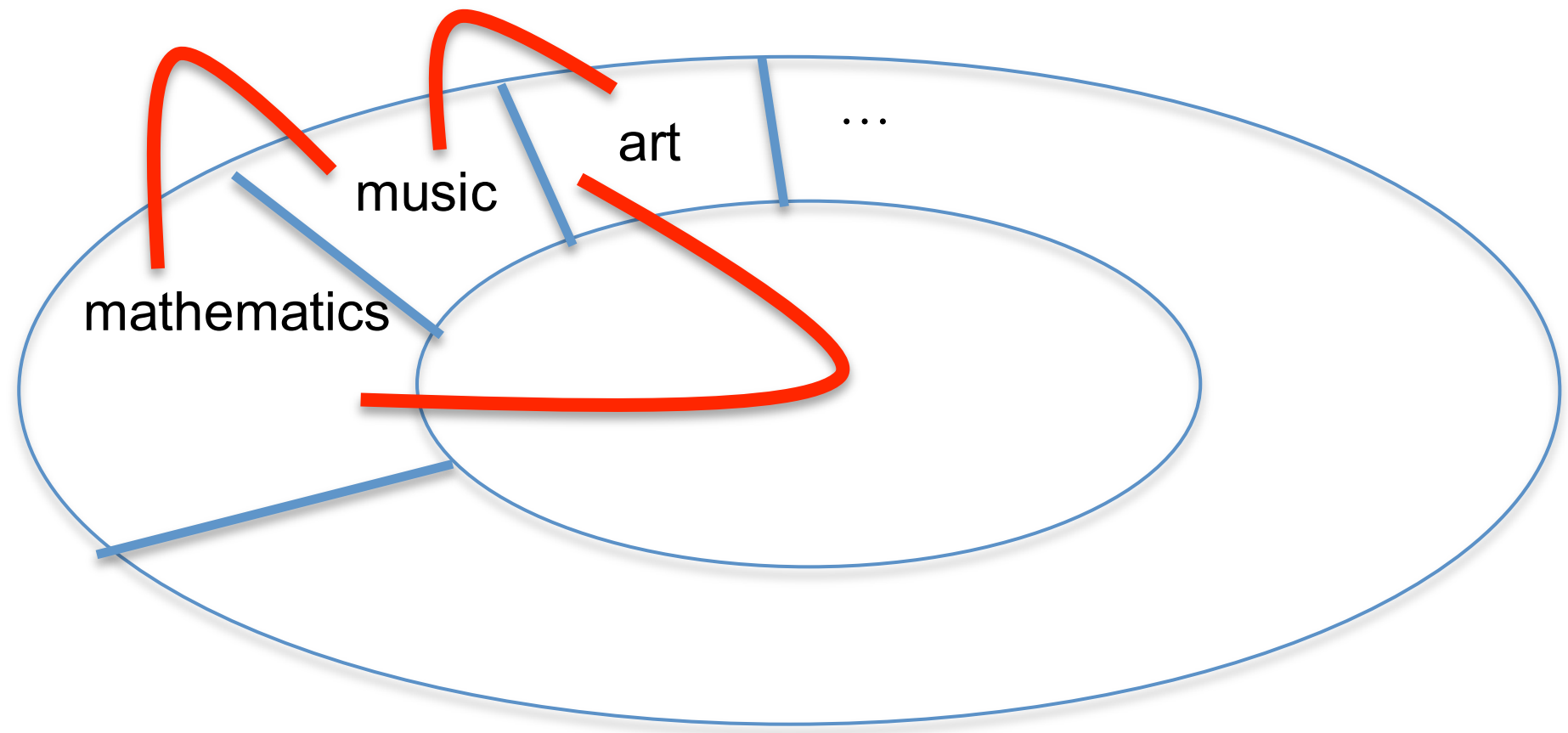

Figure 1. Schematic illustration of three mental wormholes, indicated by curved red lines, connecting concepts in one domain to concepts in another. See text for explanation.

## The Strikingly Non-compositional Manner in which Concepts and Percepts Interact

Strange things happen in mental wormholes. Concepts and percepts interact with each other in an almost chameleon-like manner, spontaneously transmuting to match their surroundings, or context. Thus a context can alters sometimes dramatically, how you experience a concept (Barsalou, 1999). Even more strangely, they often interact in ways that are non-compositional; in other words, that violate the rules of classical logic. For example, although people do not rate 'guppy' as a typical PET, nor as a typical FISH, they rate it as a highly typical PET FISH (Osherson & Smith, 1981). Something happens when the concepts PET and FISH interact that cannot be formally described using any kind of logical—or even fuzzy logical—function of the typicalities of pets and fish. Not every combination of percepts or concepts exhibits noncompositionality, but the phenomenon is widespread (Storms, De Boeck, Van Mechelen, & Ruts, 1998). Moreover, people not only prefer combinations that interact noncompositionally over those that interact in a straightforward additive manner, but find them more creative (Gibbert, Hampton, Estes, & Mazursky, 2012).

### *Emergent Properties and Incorporation of Outside Knowledge*

One spectacular manifestation of non-compositionality observed when percepts and concepts interact is the birth of new emergent properties. For example, although people do not rate 'talks' as a characteristic property of PET or BIRD, they rate it as a characteristic property of PET BIRD. 'Talks' it is not a property of either constituent concept; it is an emergent property of the combined concept. A combined concept may also exhibit a loss of properties typical of one or both constituents. For example, 'surrounded by water', a seemingly central property of ISLAND, is (hopefully) not present in instances of KITCHEN ISLAND.

A related phenomenon is that concept combinations often incorporate world knowledge that goes beyond the constituent concepts themselves (Fodor, 1998; Hampton, 1987). For example, consider the concept TEAPOT, which presumably came about by taking a known object, a pot, and adapting it to be used for liquid. The combining of these concepts gave birth to a new concept, the idea of a spout. Moreover, this process employed knowledge of gravity and fluid dynamics that extends beyond the properties and instances of TEA and POT.

### *Entanglement and Interference*

How are we to understand the interactions between percepts and concepts at the core of the creative process when we cannot even describe these interactions with standard mathematics? There is in fact a branch of mathematics that is well suited to describing highly contextual interactions. It was first used in quantum mechanics, but generalizations of these formalisms have been applied outside of physics to the description of concepts (and use of these generalizations does not imply that phenomena at the quantum level play any role). Specifically, the noncompositional manner in which concepts interact can be described using formalisms that were developed to describe the physical phenomena of entanglement and interference. The formulae for *entanglement* were developed to describe situations of non-separability where different entities form a composite entity. It uses *amplitudes,* which are similar to probabilities, except that they can exhibit wave-like interference effects. Technically, the state $|\psi\rangle_{AB}$ is separable if for the amplitudes $c_{ij}$ amplitudes $a_i$ and $b_j$ can be found such that $c_{ij} = a_i b_j$. It is inseparable, and therefore an entangled state, if this is not the case, hence if the amplitudes describing the state of the composite entity are not of a product form. Numerous studies have shown that the way in which people use and interpret concept combinations can be described using the formalisms for entanglement, and that concept combination data sets exhibit constructive and destructive interference effects (Aerts, Broekaert, Gabora, & Veloz, 2012; Aerts, Gabora & Sozzo, 2013; Aerts & Sozzo, 2011; Bruza, Kitto, Ramm, & Sitbon, 2011; Gabora & Aerts, 2002; Nelson & McEvoy, 2007).

## Conclusions and Implications for Modeling Creativity

Creative thought exhibits some rather spectacular characteristics. First, it often incorporates ingredients from beyond the domain in which it is eventually expressed (e.g., thoughts and ideas from diverse arenas of life may be woven into the creation of a piece of art or music). If anything encoded in memory can potentially be recruited into a creative thought process, the mind requires a means of (1) finding relevant material, and

(2) converting it into a form in which it is amenable to expression in a different domain. It was suggested that the domain-specific operations of a creative act operate in service of this Aware-Restructure-Express (ARE) process, which not only drives creativity in the first place but brings about potentially therapeutic cognitive restructuring.

Creative thought involves the forging of what we called mental wormholes: new associations amongst concepts and percepts that potentially come from disparate domains. This interaction may exhibit characteristics that make creativity challenging to model, such as noncompositionality, emergent properties, incorporation of outside knowledge, entanglement, and interference.

By building a model of something from scratch you can put your ideas to the test and see how it behaves under different conditions. Numerous mathematical and computational models of creativity exist (e.g., Boden, 2009; Cope, 2005; DiPaola & Gabora, 2009; Gabora & Tseng, 2014; Gero, 2000; Thagard & Stewart, 2011; Wiggins, Pearce, & Müllensiefen, 2009). With rare exceptions—e.g., some models of analogy and metaphor—these models are restricted to a single domain (e.g., a computer model of musical composition that uses only musical knowledge). A single-domain model is far easier to construct than one that incorporates multiple domains, but because it does not incorporate the weaving in of cross-domain material during the ARE process, it fails to capture not just what fuels the creative process in the first place but the potentially transformative and meaningful personal impact it may have on the creator. Therefore it may be limited with respect to what it can tell us about how the human mind creates.

If a model is to capture what is fundamental about the inner workings of human creativity it must also be able to describing the noncompositional interactions between concepts and percepts that lie at the heart of the creative act. Though fledgling efforts have been made to develop mathematical models of creativity using the quantum formalism described above (e.g., Gabora & Carbert, 2015; Veloz, Gabora, Eyjolfson, & Aerts, 2011), they have yet to be used in computer programs that generate creative outputs. We are on the way to accomplishing this, however, and a major leap forward in our understanding of creativity may lie at our doorstep.

## Acknowledgements

This research was supported in part by a grant from the Natural Sciences and Engineering Research Council of Canada.